\documentclass[prb,twocolumn,amsmath,amssymb,aps,superscriptaddress,citeautoscript,longbibliography]{revtex4-2}

\usepackage{feynmp}
\usepackage{amsmath}
\usepackage{graphicx}
\usepackage{multirow}
\usepackage{latexsym}
\usepackage{textcomp}
\usepackage{verbatim}
\usepackage{color}
\usepackage{bm}
\usepackage{subfigure}
\usepackage{everysel}
\usepackage{keyval}
\usepackage{ragged2e}
\usepackage{dsfont}
\usepackage{amssymb}
\usepackage{enumitem}
\usepackage{pstricks}
\usepackage{mathtools}

\usepackage{braket}
\usepackage{slashed} 
\usepackage{hyperref}        
\hypersetup{
     colorlinks=true,
     citecolor = cyan,    
     linkcolor=magenta,
     }

\usepackage{graphicx}
\usepackage{xcolor}
\usepackage{amsmath}
\usepackage{bbold}
\usepackage{amsfonts}
\usepackage{bbm}
\usepackage{comment}
\usepackage{lettrine}

\newcommand{\osum}{{%
    \setbox0\hbox{\circ}%
    \rlap{\hbox to \wd0{\hss\sum\hss}}\box0
}}

\begin{document}

\title{Topological magnon-photon interaction for cavity magnonics}

\author{Jongjun M. Lee}
\affiliation{Department of Physics, Pohang University of Science and Technology (POSTECH), Pohang 37673, Korea}

\author{Myung-Joong Hwang}
\email{Electronic Address: myungjoong.hwang@duke.edu}
\affiliation{Division of Natural and Applied Sciences, Duke Kunshan University, Kunshan, Jiangsu 215300, China}
\affiliation{Zu Chongzhi Center for Mathematics and Computational Science, Duke Kunshan University, Kunshan, Jiangsu 215300, China}

\author{Hyun-Woo Lee}
\email{Electronic Address: hwl@postech.ac.kr}
\affiliation{Department of Physics, Pohang University of Science and Technology (POSTECH), Pohang 37673, Korea}

\begin{abstract}
\begin{center}
    \textbf{Abstract}
\end{center}
The study of cavity magnonics and topological insulators has made significant advances over the past decade, however the possibility of combining the two fields is still unexplored. Here, we explore such connection by investigating hybrid cavity systems that incorporate both a ferromagnet and a topological insulator. We find that electrons in the topological surface state efficiently mediate the effective electric dipole coupling between the spin of the ferromagnet and the electric field of the cavity, in contrast with the conventional cavity magnonics theory based on magnetic dipole coupling. We refer to this coupling as topological magnon-photon interaction, estimating it one order of magnitude stronger than the conventional magnon-photon coupling, and showing that its sign can be manipulated. We discuss the potential of our proposed device to allow for scaling down and controlling the cavity system using electronics. Our results provide solid ground for exploring the functionalities enabled by merging cavity magnonics with topological insulators.
\end{abstract}

\date{\today}
\maketitle

\noindent{{\textbf{Introduction}}}\\The magnon collective excitation of the spin has been recently introduced in cavity quantum electrodynamics (cQED) systems called the cavity magnonics \cite{lachance2019hybrid,li2020hybrid,rameshti2021cavity, pirro2021advances, yuan2022quantum}. In this research field, the magnetic system coupled to the cavity provides physical degrees of freedom that are not typically found in a traditional cQED system, and its unique non-reciprocal properties offer opportunities for its application in various fields \cite{simon2007single,tanji2009heralded,zhang2015magnon,hisatomi2016bidirectional,osada2016cavity, zhang2016optomagnonic,lachance2020entanglement}. The cavity magnonics can be used to generate quantum states of magnets that are useful for quantum information processing; for example, the magnon naturally can have the bosonic squeezing without the parametric driving \cite{kamra2019antiferromagnetic, kamra2020magnon, sharma2021spin} and the antibunching intrinsically \cite{yuan2020magnon}. It can also realize a non-Hermitian system with the (anti-)parity-time symmetric Hamiltonian by inducing the dissipative magnon-photon coupling inside the cavity \cite{harder2018level, zhang2019higher, yao2019microscopic, yu2019prediction, wang2019nonreciprocity, yang2020unconventional, wang2020dissipative, yuan2020steady}. 

Strong magnon-photon coupling can provide opportunities for studying and controlling the cavity magnonics system, including the above. However, achieving the strong coupling regime in current studies requires a sizable magnet volume since the present strategy to achieve the strong coupling regime relies on the proportionality of the coupling strength to $\sqrt{N}$ where $N$ is the number of spins in the magnet \cite{huebl2013high, zhang2014strongly, tabuchi2014hybridizing, goryachev2014high,bourhill2016ultrahigh}. Unfortunately, this sizable volume can introduce quantum decoherence through the inhomogeneous broadening since achieving the homogeneity within a large volume becomes increasingly difficult due to electromagnetic field variations, material defects and other geometric factors \cite{houdre1996vacuum,chotorlishvili2011two,lodahl2015interfacing,trivedi2019photon}. Therefore, exploring alternative methods to enhance the coupling strength beyond the current mechanism is necessary. Achieving this goal could even lead to the realization of the ultrastrong coupling regime with a reasonable number of spins, which would be a significant advancement in the field of cavity magnonics.

\begin{figure}[t]
    \centering
    \includegraphics[width=8cm]{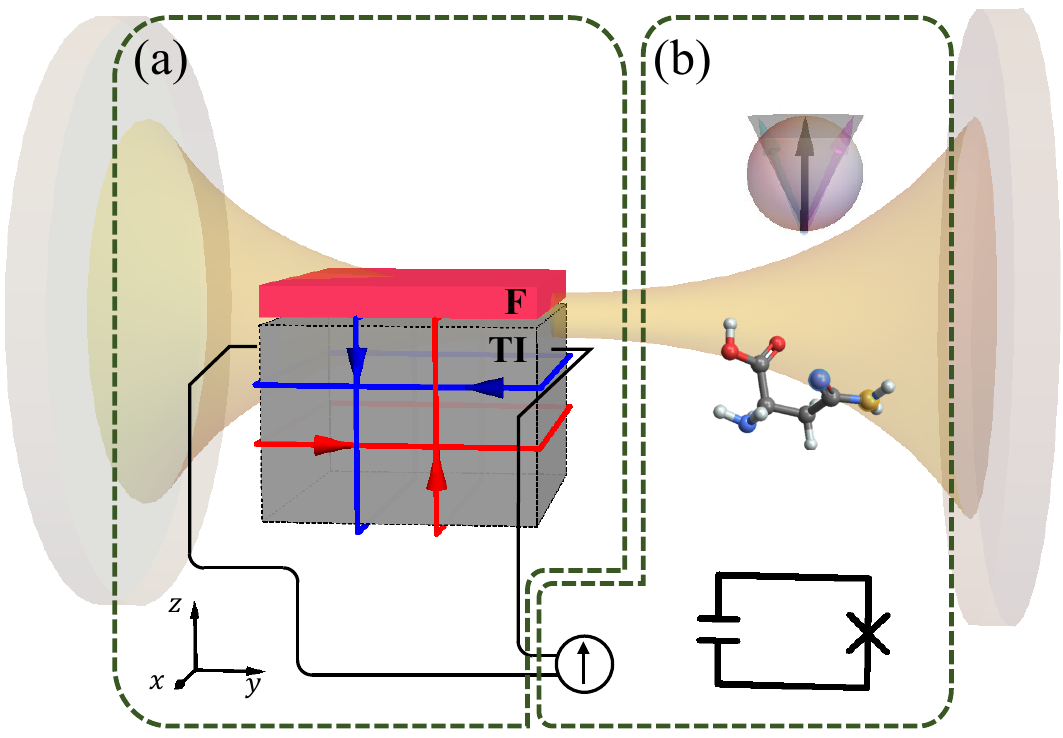}
    \caption{Schematic illustration of the setup. \textbf{(a)} A heterostructure of the ferromagnet (F, pink) and the topological insulator (TI) is set inside the cavity. TI's surface states with different chiralities are represented by arrows in distinct colors, red and blue. \textbf{(b)} The topological insulator is connected to the current source. And other physical objects such as magnets, molecules and qubits are placed inside the cavity.}
    \label{FIG_1}
\end{figure}

In this work, we propose a mechanism for the magnon-photon interaction. We introduce a heterostructure of a topological insulator and a ferromagnet in the cavity magnonics system (FIG.~\ref{FIG_1}) \cite{qi2008topological,hasan2010colloquium}. The topological insulator has been traditionally used in the spintronics to strongly couple the spin degrees of freedom to an electric field, providing a strong magnetoelectric effect and spin-orbit torque \cite{pesin2012spintronics,vsmejkal2018topological,breunig2022opportunities,he2022topological}. We show that electrons in the topological surface state of the topological insulator can mediate an effective coupling between the magnetization of the ferromagnet and the cavity electric field. Thus, the resulting effective interaction between magnons and photons relies on the electric dipole coupling, and is fundamentally different from the conventional magnon-photon coupling that relies on the magnetic dipole coupling. The effective magnon-photon coupling, which we call the topological magnon-photon interaction, is also proportional to $\sqrt{N}$, similar to the conventional coupling. For given $N$, we estimate that the effective magnon-photon coupling is one order of magnitude stronger than the conventional magnon-photon coupling.

The sign of the topological interaction depends on the relative sign between the chirality of the surface and the magnetization. We design a setup where the sign of the interaction is a gauge invariant quantity and generates measurable consequences. We propose a method to control the cavity magnonics system using the electric current through the topological insulator. The two-dimensional setup of the cavity magnonics is also proposed using the van der Waals material and the two-dimensional cavity. This compact device allows the electrical control of the quantum states of the magnon.

\vspace{5mm}

\noindent{\textbf{Results}}\\
{\small \textbf{Model And Setup.}} This study focuses on a cavity system composed of a topological insulator and an insulating ferromagnet. The ferromagnet is stacked on top of the topological insulator, creating a heterostructure as depicted in FIG.~\ref{FIG_1}(a). The total Hamiltonian of the cavity system consists of six parts describing photons, magnons, and electrons together with their mutual interactions. The cavity contains an electromagnetic field with an energy scale ranging from microwave to far-infrared wavelengths, and its classical Hamiltonian density has the form $\mathcal{H}_{ph} = \epsilon \bm{E}^{2}/2+\bm{B}^{2}/2\mu $. This can be quantized into the harmonic oscillator Hamiltonian of the photon operator with the frequency $\omega= c|\bm{k}|$.
\begin{equation}
    H_{\text{ph}} = \sum_{\bm{k}} \hbar \omega_{\bm{k}} \Big[a^{\dagger}_{\bm{k}}a_{\bm{k}} + \frac{1}{2}\Big],
\end{equation}
where $a_{\bm{k}}$ and $a^{\dagger}_{\bm{k}}$ are the photon operator with momentum $\bm{k}$. The ferromagnet is described by the spin Hamiltonian $H_{\text{s}} = -J_{\text{ex}} \sum_{\langle i j \rangle} \bm{S}_{i}\cdot \bm{S}_{j} - K\sum_{i} S^{z2}_{i} + \cdots$ where we have taken into account the exchange coupling, the anisotropy, and other contributions. The magnetization is assumed to be aligned along $\hat{z}$ that is out of the plane of the topological insulator's surface. With the Holstein-Primakoff transformation, the general spin Hamiltonian can be expressed as a sum of squeezed harmonic oscillators \cite{holstein1940field}. 
\begin{equation} 
    H_{\text{mag}}/\hbar= \sum_{\bm{k}} ( \omega^{\text{m}}_{\bm{k}}  m^{\dagger}_{\bm{k}}m_{\bm{k}} + \alpha_{\bm{k}} m^{\dagger}_{\bm{k}}m^{\dagger}_{-\bm{k}} + \alpha^{*}_{\bm{k}} m_{\bm{k}}m_{-\bm{k}}  ),
\end{equation}
where $m_{\bm{k}}$ and $m^{\dagger}_{\bm{k}}$ are the magnon operator with momentum $\bm{k}$. The magnon frequency $\omega^{\text{m}}_{\bm{k}}$ depends on the microscopic detail of the spin Hamiltonian, and we ignore the squeezing terms by considering only the uniaxial anisotropy and the external field parallel to the ordering for simplicity. We here consider only the topological surface state of the topological insulator interacting with the insulating ferromagnet at the interface. Its bare theory is described by the massless Dirac electron with the chirality, i.e.
\begin{equation}
    H_{0} = v_{\text{F}} \int d^{2}\bm{r}\: \psi^{\dagger}_{\bm{r}}  \bm{\sigma} \cdot (-i\hbar \bm{\nabla} \times \hat{z}) \psi_{\bm{r}},
\end{equation}
where $\psi_{\bm{r}}= (\psi_{\bm{r}\uparrow},\psi_{\bm{r}\downarrow})$ is the spinor basis at the position $\bm{r}$ and $v_{\text{F}}$ is the Fermi velocity of the surface state \cite{fu2007topological,hasan2010colloquium}. Note that the spin vector is aligned perpendicular to the momentum vector to form a chiral texture on the momentum space plane in the ground state.

Photons, magnons, and electrons interact with each other. First, the magnetic field of the cavity and the spins of the ferromagnet directly interact with each other through the magnetic dipole interaction. The classical form of the Hamiltonian is $H_{\text{Z}} = - \int d\bm{r} \: \bm{m} \cdot \bm{B}$ where $\bm{m}=\gamma \bm{S}$ is the magnetization of the ferromagnet and $\gamma$ is the gyromagnetic ratio. By quantizing the fields, we obtain the conventional magnon-photon interaction that reads
\begin{equation}
H_{\text{Z}}/\hbar = - g_{\text{m}}  (a-a^{\dagger})(m-m^{\dagger}), 
\label{Eq_Zeeman_MPh}
\end{equation}
with the magnon-photon coupling strength
\begin{equation}
    g_{\text{m}} = \frac{\gamma }{c} \eta \sqrt{\frac{E_{\text{g}}NS }{2\epsilon V_{\text{m}}}},
\end{equation}
where $V_{\text{m}}$ is the magnet volume, $\eta$ is the magnon filling factor, $S$ is the dimensionless spin amplitude, $\epsilon$ is the permittivity, and $E_{\text{g}}=\hbar \omega^{\text{m}}_{0}$ is the magnon gap energy \cite{soykal2010strong,soykal2010size,goryachev2014high,bourhill2016ultrahigh,goryachev2018cavity,flower2019experimental}. The cavity magnetic field is assumed to be in the $y$ direction. $a$ and $m$ are the photon and the magnon operator of the uniform mode, respectively. Note that the coupling strength is enhanced by $\sqrt{N}$, where $N$ is the number of the total spins in the system.

The electrons at the surface state of the topological insulator interact with the electric field of the cavity through the electric dipole interaction, which can be described by the minimal coupling $-i\hbar\bm{\nabla} \rightarrow (-i\hbar \bm{\nabla} -e\bm{A})$ where $\bm{A}$ is the gauge field of the cavity. In addition, the electron interacts with the magnetic moment of the ferromagnet through the exchange coupling. There is also a direct interaction between the electron spins in the topological insulator and the magnetic field of the cavity. However, this interaction is ignored due to its negligible coupling strength compared to the exchange coupling. Therefore, the Hamiltonian density of the topological surface state is given as
\begin{equation}
    \mathcal{H}_{\text{top}} = v_{\text{F}} \bm{\sigma}\cdot [(-i\hbar \bm{\nabla} - e\bm{A}) \times \hat{z}] - J \bm{\sigma }\cdot \bm{m}.
\label{Eq_H_top}
\end{equation}
The magnetization in the ferromagnet is $\bm{m}=m_{x}\hat{x}+m_{y}\hat{y}+M\hat{z}$ where the magnetization is aligned along $\hat{z}$, i.e., $M\gg m_{x},\:m_{y}$, and it will be quantized into the magnon operator later. The last term is the exchange coupling which is ferromagnetic, i.e., $J>0$, between the electron spin in the surface state and the magnetic moment in the ferromagnet. For notational simplicity, the spinor basis was rotated by $\pi/2$ along $\sigma_z$ \cite{garate2010inverse}. Then, one can rewrite the Hamiltonian density as
\begin{equation}
    \mathcal{H}'_{\text{top}} = v_{\text{F}}\sum_{\alpha=x,y}(-i\hbar\partial_{\alpha}-e\tilde{A}_{\alpha})\sigma_{\alpha} - JM \sigma_{z},
\end{equation}
which is the $\text{QED}_{3}$ theory of the massive chiral Dirac electron \cite{peskin2018introduction}. The exchange coupling terms in $\hat{x}$ and $\hat{y}$ are absorbed to the minimal coupling by redefining the gauge field as $\tilde{A}_{x}= A_{x}-Jm_{y}/ev_{\text{F}}$ and $\tilde{A}_{y} = A_{y}+Jm_{x}/ev_{\text{F}}$.

\vspace{5mm}

{\small \noindent{\textbf{Calculation of the magnon-photon interaction.}}} Our goal in this section is to calculate the total magnon-photon interaction in this model. In addition to the direct interaction $H_{\text{Z}}$ in Eq.~(\ref{Eq_Zeeman_MPh}), there appears the topological interaction mediated by the electrons in the topological surface state.

To calculate the topological magnon-photon interaction, we integrate out the electronic degrees of freedom. We take the path integral formalism to perform the integration. The action $S_{\text{top}}$ of the topological surface state and its interaction with the cavity field and the ferromagnet defines the following effective action $S_{\text{eff}}$ after integrating out the electron field,
\begin{equation}
\begin{aligned}
    e^{-S_{\text{eff}}} =  \int D\bar{\psi}D\psi \: e^{-S_{\text{top}}},
\end{aligned}
\end{equation}
where $S_{\text{top}} = \int dt d\bm{r} \: \bar{\psi} (\partial_0 - \mu - \mathcal{H}'_{\text{top}}) \psi$ and $\mu$ is the chemical potential. The integration is perturbatively calculated up to the one-loop level of the electron since the electron-photon interaction strength determined by the fine-structure constant is small \cite{peskin2018introduction}. This way, one obtains the effective action $S_{\text{eff}}=\int dt d^{2}\bm{r} \mathcal{L}_{\text{eff}}$ with the Lagrangian density
\begin{equation}
    \mathcal{L}_{\text{eff}} \simeq \frac{e^{2}}{24\pi J|M|} \tilde{F}_{\mu\nu}\tilde{F}^{\mu\nu} - \frac{e^{2}c}{4\pi h} \frac{M}{|M|} \epsilon^{\mu\nu\rho} \tilde{A}_{\mu}\partial_{\nu}\tilde{A}_{\rho},
    \label{Eq_eff_L1}
\end{equation}
where $\epsilon^{\mu\nu\rho}$ is the Levi-Civita symbol and $\tilde{F}_{\mu\nu}= \partial_{\mu}\tilde{A}_{\nu}-\partial_{\nu}\tilde{A}_{\mu}$ is the redefined electromagnetic field tensor \cite{babu1987derivative,deser1988c,dunne1999aspects}. Since $\tilde{F}_{\mu\nu}$ is gauge invariant, it is evident that the first term of $\mathcal{L}_{\text{eff}}$ is gauge invariant. On the other hand, the second term is not expressed in terms of the field tensor and its gauge invariance is not obvious. However, it is also gauge invariant since its structure agrees with that of the Chern-Simons term that is gauge invariant \cite{dunne1999aspects,dunne1990topological,tong2018gauge}. Physically, the second term, which we call the Chern-Simons term, arises from the chiral Dirac electron of the topological surface state.

$\mathcal{L}_{\text{eff}}$ contains cross terms of the gauge field $A_{\mu}$ and the magnetization $m_{\nu}$ since $\tilde{A}_{\mu}= A_{\mu}+ J \epsilon^{\mu z \nu }m_{\nu}/ev_{\text{F}}$. These cross terms amount to the indirect interaction between the spin of the ferromagnet and the electric field of the cavity. One can derive the indirect magnon-photon interaction by quantizing the cross terms. The cross terms from the first term of $\mathcal{L}_{\text{eff}}$ are inversely proportional to the effective mass $JM$ and are irrelevant in the renormalization group sense \cite{dunne1999aspects}. In contrast, the Chern-Simons term is not inversely proportional to the effective mass $JM$. Thus, the Chern-Simons term is more relevant than the first term in the low-temperature limit, and we only keep the cross terms from the Chern-Simons term. The effective Lagrangian density becomes
\begin{equation}
    \begin{aligned}
    \mathcal{L}_{\text{eff}} \simeq \mathcal{L}^{0}_{\text{CS}} - \frac{J ec}{2\pi v_{\text{F}} h} \frac{M}{|M|} (A_{y}\partial_0 m_{y}+A_{x}\partial_0 m_{x}),
    \label{Leff1}
    \end{aligned}
\end{equation}
where $\mathcal{L}^{0}_{\text{CS}}$ is the conventional Chern-Simons term (or the second term of Eq.~(\ref{Eq_eff_L1}) in the limit $J=0$). The terms with two magnetization components are neglected since they give negligible correction of the magnon frequency. Next, the electromagnetic field and the magnetization can be quantized into bosonic operators. The electric field in the cavity is assumed to be along $\hat{x}$. Their uniform modes couple each other since the magnetization feels a nearly uniform electromagnetic field due to the longer wavelength of the light than the spin wave. Now we have the cavity magnonics Hamiltonian (see methods for calculation details),
\begin{equation}
    H_{\text{mp}}/\hbar = \Omega m^{\dagger}m + \omega a^{\dagger}a + ig_{\text{e}} (m+m^{\dagger})(a-a^{\dagger}),  
    \label{H_mag-ph_1}
\end{equation}
with the indirect magnon-photon coupling strength mediated by the topological surface state,
\begin{equation}
    g_{\text{e}} = - \frac{J A e }{v_{\text{F}} h^{2}}\frac{M}{|M|}\eta \sqrt{\frac{E_{\text{g}}NS}{2\epsilon V_{\text{m}}}},
    \label{H_mag-ph_2}
\end{equation} 
where $A$ is the unit cell area, $\eta$ is the magnon filling factor, and $V_{\text{m}}$ is the magnet volume \cite{bourhill2016ultrahigh,flower2019experimental}.

Equations~(\ref{H_mag-ph_1}) and (\ref{H_mag-ph_2}) are our main results. A few remarks are in order. The first remark is that the magnon and the cavity photon are electrically coupled in the theory. In the conventional cavity magnonics system, the magnon-photon coupling originates from the Zeeman interaction $H_{\text{Z}}$ which is the magnetic dipole interaction between the spin and the magnetic field \cite{soykal2010strong,soykal2010size}. Since neither the magnon nor the spin has a charge, the electric dipole coupling is typically absent. However, the topological surface state can glue the magnon and the electric field of the cavity photon since the electrons in the topological surface state couple to the electric field through the electric dipole interaction, and the electrons couple to the magnon or the spin through the spin-momentum coupling. These two steps are mechanisms of our magnon-photon interaction. The second remark is the $\sqrt{N}$ dependence of the coupling strength [Eq.~(\ref{H_mag-ph_2})], implying that the coupling strength can be enhanced by increasing the spin density and the system size, which is similar to the conventional magnon-photon coupling through the magnetic dipole interaction \cite{soykal2010strong,soykal2010size}. The final remark is the universality of our theory. It only requires the chiral spin-momentum locking in the gapped topological surface state but does not depends on the details such as the magnon band and the material details. 

\vspace{5mm}

\noindent{\textbf{Discussion.}}\\
To quantitatively compare the topological and conventional coupling strengths, we estimate the ratio 
\begin{equation}
    \frac{g_{\text{e}}}{g_{\text{m}}} = \frac{JAe c}{ \gamma v_{\text{F}} h^{2} },
\end{equation}   
for the same values of the magnon filling factor, which is independent of $N$. We used the value $5\cdot 10^{5}$ $\text{m s}^{-1}$ for the fermi velocity $v_{\text{F}}$ of the topological surface state \cite{hsieh2008topological,xia2009observation,chen2009experimental,zhang2009topological} and $0.3$ eV for the exchange coupling strength $J$ between the spin in the state and the magnetic moment in the ferromagnet \cite{liu2009magnetic}. Our estimation reveals that the topological magnon-photon interaction is expected to be at least about 15 times stronger than the conventional magnon-photon interaction with the same filling factor value. This demonstrates how $g_{\text{e}}$ can achieve the desired coupling strength with a smaller-sized (smaller $N$) magnet, which can help minimize quantum decoherence in the cavity magnonics system. In larger magnets, higher-order magnon modes can be excited more easily due to the inhomogeneity of the cavity field, which can result in extra quantum decoherences of the magnon. Hence, reducing the magnet's size can be advantageous for decreasing decoherence in the cavity magnonics system \cite{houdre1996vacuum,chotorlishvili2011two,lodahl2015interfacing,trivedi2019photon,lachance2019hybrid}.

\begin{figure}[t]
    \centering
    \includegraphics[width=8.5cm]{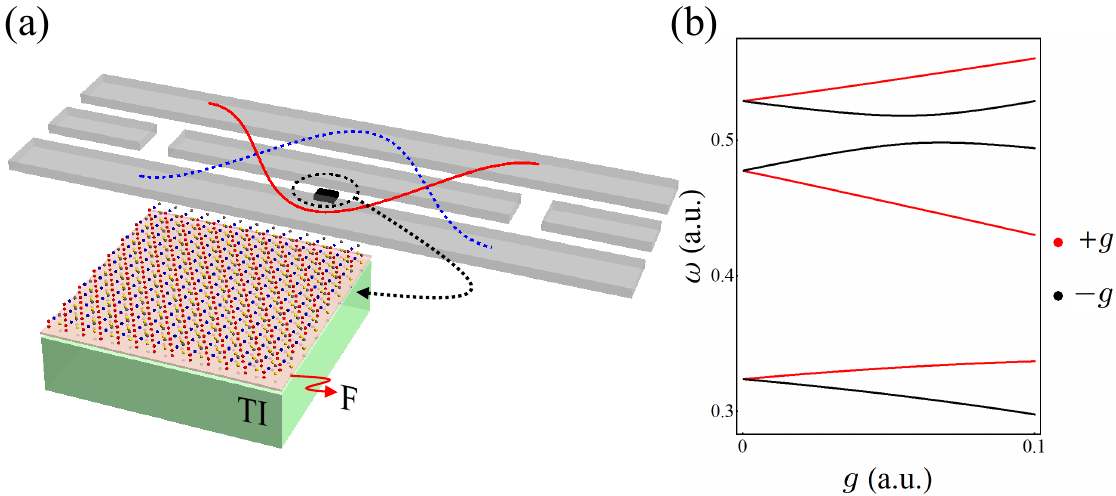}
    \caption{Experimental setup for two-dimensional cavity magnonics and energy level spectra of a hybrid system. \textbf{(a)} Illustration of the setup, where a layer of two-dimensional magnetic material (F, pink) is placed on top of a topological insulator (TI, green) within a two-dimensional cavity. The curves represent the electromagnetic field within the cavity. \textbf{(b)} Energy level spectra of the hybrid system $\omega$, as described by the Hamiltonian in Eq.~(\ref{Eq_Nonlinear}), with the coupling strength changing sign from $g\text{ (red)}$ to $-g\text{ (black)}$.}
    \label{FIG_2}
\end{figure}

As a concrete example, we propose a cavity magnonics setup with a heterostructure between a two-dimensional magnetic material and a topological insulator as illustrated in FIG.~\ref{FIG_2}(a). There are a few van der Waals materials known to be electrically insulating and exhibit two-dimensional magnetic ordering, which is attributed to their strong spin-orbit coupling and the Coulomb interaction: $\text{CrI}_{3}$ \cite{lado2017origin,huang2018electrical,cenker2021direct}, $\text{Cr}_{2}\text{Ge}_{2}\text{Te}_{6}$ \cite{han2019topological,zhuo2021manipulating} to name a few. Using two-dimensional materials is expected to have several advantages. First, its dimensionality basically guarantees the integration of the device in analogy to the atomically thin monolayer laser with the transition-metal dichalcogenides \cite{ye2015monolayer}. Also, one can engineer the magnonic properties of the van der Waals magnet in many different ways with simple manipulations, for example, stacking, twisting, applying strain, etc. The van der Waals magnet does not only provide a wide range of the magnon frequency from GHz to THz, but it also allows tuning of the magnon frequency by using the external field \cite{yuan2021electrically,xiao2022tunable}. It is known that the magnon dissipation rate of the van der Waals material is usually larger than the one of the bulk magnets; however, there are few studies on the magnon dissipation rate in the van der Waals material. Also, many of them have low critical temperatures in the range of a few tens of Kelvin, although few of them, such as $\text{Fe}_{5}\text{Ge}\text{Te}_{2}$, have higher critical temperatures of around the room temperature \cite{zhuo2021manipulating,cenker2021direct,may2019ferromagnetism}. Nevertheless, various two-dimensional ferromagnetic materials are actively being discovered and studied currently. Additionally, we expect the use of two-dimensional cavities, such as plasmon cavities \cite{kaliteevski2007tamm,lheureux2015polarization,messelot2020tamm}, photonic crystals \cite{vuvckovic2001design,zhang2016collective,sengupta2018terahertz,yang2020terahertz}, meta-materials \cite{scalari2012ultrastrong,bayer2017terahertz} and microwave resonators \cite{hou2019strong,mandal2020coplanar,zhang2021zero}. The microwave resonator, consisting of a superconductor, necessitates operation at cryogenic temperatures. They can provide benefits in terms of compactness and high quality factor of the cavity.

We perform a quantitative estimation of the magnon-photon coupling strength and compare it with the experimental value of the conventional mechanism. To make a comparison, we define a quantity $\tilde{g}_{\text{e/m}}=g_{\text{e/m}}/\sqrt{N}$ which represents the magnon-photon interaction per spin, since the coupling strength is enhanced by the number of spins in the magnet. Our estimation reveals that for YIG, the coupling strengths are $\tilde{g}_{\text{m}}\simeq 0.21 - 2.03$ Hz for the conventional mechanism and $\tilde{g}_{\text{e}}\simeq 3.18 - 30.3$ Hz for our topological magnon-photon interaction. It is worth noting that the conventional cavity magnonics experiments have observed a value $\tilde{g}_{\text{m}}\simeq 0.21-2.66$ Hz under the same magnon filling factor for YIG and LiFe in both cryogenic and ambient temperatures \cite{huebl2013high,zhang2014strongly,tabuchi2014hybridizing,goryachev2014high,bourhill2016ultrahigh}.
We also estimate the value $\tilde{g}_{\text{e}}\simeq 1.01-9.64$ Hz for the $\text{CrI}_{3}$ ferromagnetic monolayer under the same geometric conditions of the cavity \cite{cenker2021direct}. The results demonstrate that our theory predicts a larger magnon-photon interaction strength per spin compared to the ordinary magnetic dipole interaction. In our calculations, we used values of the exchange coupling strength 0.3 eV for YIG \cite{liu2009magnetic}, and 0.123 eV for $\text{CrI}_{3}$ \cite{su2022magnetic}; however, further studies are necessary to clarify the strength for the van der Waals material. And, the magnetic moments $\mu_{\text{Cr}^{3+}}\simeq 3\mu_{\text{B}}$ and $\mu_{\text{Fe}^{3+}}\simeq 5\mu_{\text{B}}$ were used in the calculation \cite{tang2020magnetic,dolui2020proximity,goryachev2018cavity}. The magnon frequency was measured at the $\Gamma$ point of the magnon dispersion without the external magnetic field.


One intriguing feature of the topological magnon-photon interaction in Eq.~(\ref{H_mag-ph_1}) is the dependence of its sign on the relative orientation of the chiral spin-momentum texture of the surface and the magnetization of the ferromagnet. This sign change does not typically affect the physics with a single magnet in the cavity since it is nothing but the change of the global gauge choice. However, the sign is no longer gauge degrees of freedom and becomes relevant if there are other interacting degrees of freedom. For example, one can consider additional modes in the hybrid system such as the photon of the multi-mode cavity and the phonon of the mechanical oscillator. The additional modes may be described by the following Hamiltonian. 
\begin{equation}
\begin{aligned}
    H_{l}/\hbar &= \sum_{j=1,2,3} \omega_{j} a^{\dagger}_{j}a_{j} + \sum_{j<l}g_{jl} (a_{j}+a^{\dagger}_{j})(a_{l}+a^{\dagger}_{l}),
\end{aligned}
\label{Eq_Nonlinear}
\end{equation}
where, for example, $a_{1}$ is the magnon, $a_{2}$ is the cavity photon, $a_{3}$ is the additional mode and $g_{12}=g$. The additional interaction could be extended into the dissipative coupling as well as the coherent coupling in general. Note that the sign change $g\rightarrow -g$ modifies the energy level as illustrated in FIG.~\ref{FIG_2}(b). Thus the sign change is no longer the gauge transformation in this Hamiltonian. We suggest that this property could provide an attractive opportunity to control the cavity magnonics system. First, one can consider an electronic-circuit integrated setup of the cavity magnonics illustrated in FIG.~\ref{FIG_1}(a) and (b). The topological insulator is connected to the current source, and additional devices such as other magnetic materials, molecules or qubits that interact with the magnet or cavity field can be placed within the cavity. As briefly mentioned in the introduction, the topological insulator has a large spin-orbit torque; thus the electric current can switch the magnetization in the topological insulator-ferromagnet heterostructure \cite{fan2014magnetization,han2017room,wang2017room,dziom2017observation}. Since the magnetization switching changes the sign of the magnon-photon interaction, one could switch the sign of the interaction electrically on the electronic circuit. If the sign is a gauge invariant quantity in the system as the above Hamiltonian, the electric control of the sign could manipulate the physical properties of the system such as the energy level. Additionally, the magnetization switching can shift the magnon frequency under the external magnetic field whether the interaction term's sign change is gauge invariant or not. The frequency shift can induce a mismatch of the frequencies between the cavity photon and the magnon that can weaken the magnon-photon coupling strength. Thus, one could couple or decouple magnonic devices inside the cavity system using the electronic current. These ideas should open ways for fast and in-situ control of the quantum devices in the cavity system.

\vspace{5mm}

\noindent{\textbf{Conclusion}}\\
In summary, this paper presents a topological magnon-photon interaction mediated by the topological insulator. The Chern-Simons term of the redefined gauge field has been derived as an effective field theory by integrating out the electron field. The term generates an interaction between the ferromagnet's spin and the electric field of the cavity, which follows a conventional form of the cavity magnonics Hamiltonian. The coupling strength per spin is demonstrated to be stronger than the ordinary coupling by one order of magnitude. To provide estimates of the enhancement, we evaluated the coupling strengths per spin for van der Waals magnetic materials using realistic data. Furthermore, the sign of the magnon-photon coupling depends on the relative direction between the magnetization and the electron's chirality. This property could offer an avenue for manipulating the cQED system using electric current with multiple magnets or other devices. Overall, this study bridges the gap between topological matter and cQED.

\vspace{5mm}

\noindent{\textbf{Methods}}\\
{\small \textbf{Detailed derivation of the magnon-photon coupling.} One can rewrite Lagrangian terms of the interaction from the effective Lagrangian density in Eq.~(\ref{Leff1}) by using the integration by parts.  
\begin{equation}
    \begin{aligned}
        L'_{\text{int}} &=  \frac{Je}{2\pi v_{\text{F}} h} \frac{M}{|M|} \int d\bm{r} \sum_{j=x,y} E_{j}(\bm{r},t) m_{j} (\bm{r},t),
    \end{aligned}
\end{equation}
where $E_{j}=c\partial_{0}A_{j}$ is the electric field of the cavity and $m_{j}$ is the magnetization of the ferromagnet. One should quantize the fields into the operator form to calculate the magnon-photon interaction. The electric field is
\begin{equation}
    \begin{aligned}
        \bm{E}(\bm{r},t) &= \sum_{\bm{k}} i \sqrt{\frac{\hbar \omega_{\bm{k}}}{2\epsilon V}}
        \Big[a_{\bm{k}}e^{i(\bm{k}\cdot\bm{r}-\omega_{\bm{k}}t)}  - \text{H.c.} \Big]\hat{x},
    \end{aligned}
\end{equation}
where $a_{\bm{k}}$ and $a^{\dagger}_{\bm{k}}$ are the photon operator, $V$ is the cavity volume, and $\epsilon$ is the permittivity. The electric field is assumed to be along the $\hat{x}$. The magnetization is
\begin{equation}
\begin{aligned}
    m_{x} (\bm{r},t) &= \sqrt{\frac{2S}{N}} \sum_{\bm{k}}\Big[ m_{\bm{k}} e^{i(\bm{k}\cdot\bm{r}-\omega^{\text{m}}_{\bm{k}}t)} + \text{H.c.} \Big],\\
    m_{y} (\bm{r},t) &= -i\sqrt{\frac{2S}{N}} \sum_{\bm{k}}\Big[ m_{\bm{k}} e^{i(\bm{k}\cdot\bm{r}-\omega^{\text{m}}_{\bm{k}}t)} - \text{H.c.} \Big],
\end{aligned}
\end{equation} 
where $m_{\bm{k}}$ and $m^{\dagger}_{\bm{k}}$ are the magnon operator, and $S$ is the dimensionless spin amplitude. Since the electromagnetic field is much faster than the spin wave, the electromagnetic field is almost spatially uniform for the spin wave. Also, the cavity frequency is tuned to the magnon gap frequency for the resonance. 
\begin{equation}
    \bm{E}(t) =   i \sqrt{\frac{E_{\text{g}}}{2\epsilon V}}
        \Big[a e^{-iE_{\text{g}}t/\hbar}  - \text{H.c.} \Big]\hat{x},
\end{equation}
where $E_{\text{g}} = \hbar \omega^{\text{m}}_{\bm{k}=0}$ is the magnon gap energy and $a_{\bm{k}=0}=a$. Inserting the electric field and the magnetization, the Lagrangian becomes 
\begin{equation}
    \begin{aligned}
        L'_{\text{int}} &=  \frac{iJe}{4\pi v_{\text{F}} h} \frac{M}{|M|} \sqrt{\frac{E_{\text{g}}S}{\epsilon VN}} \int d\bm{r} 
        \sum_{\bm{k}}\Big[a e^{-iE_{\text{g}}t/\hbar} \\& - \text{H.c.} \Big]  
        \Big[ m_{\bm{k}} e^{i(\bm{k}\cdot\bm{r}-\omega^{\text{m}}_{\bm{k}}t)} + \text{H.c.} \Big] .
    \end{aligned}
\end{equation}
The integral calculation shows that only the uniform mode of the magnon, the Kittel mode, couples to the electric field. By considering the geometrical filling of the magnet inside the cavity \cite{flower2019experimental}, one can calculate the magnon-photon interaction of the Hamiltonian $H'_{\text{int}}/\hbar = ig_{\text{e}} (a  - a^{\dagger} ) ( m  + m^{\dagger} )$ with $m_{\bm{k}=0}= m$ in the Schrodinger picture and the magnon-photon coupling strength,
\begin{equation}
    g_{\text{e}}= -\frac{JAe}{v_{\text{F}} h^{2}} \frac{M}{|M|} \sqrt{\frac{E_{\text{g}}NS}{2\epsilon V_{\text{m}}}},
\end{equation} 
with the magnon filling factor,
\begin{equation}
    \eta = \sqrt{ \frac{ (\int_{V_{\text{m}}} d\bm{r}\: E_{x}(\bm{r}) )^{2} }{V_{\text{m}} \int_{V} d\bm{r}\: E^{2}_{x}(\bm{r}) } },
\end{equation}
where $V_{\text{m}}$ is the magnet volume. The result is corresponding to Eq.~(\ref{H_mag-ph_1}) and Eq.~(\ref{H_mag-ph_2}).

}

\vspace{5mm}
\noindent{\textbf{Data availability}}\\
{\small Data in the manuscript can be provided from the corresponding author on request.
}

\vspace{5mm}
\noindent{\textbf{Acknowledgements}}\\
{\small J.M.L. and H.-W.L. acknowledge support from the National Research Foundation of Korea (NRF) grant funded by the Korea government (MSIT) (Grant No. 2020R1A2C2013484) and the Samsung Science and Technology Foundation (BA-1501-51). J.M.L. was supported by the POSCO Science Fellowship of the POSCO TJ Park Foundation. M.-J.H. was supported by the Startup Fund from Duke Kunshan University, and Innovation Program for Quantum Science and Technology 2021ZD0301602. J.M.L. thanks Sebae Park, Gwan-Suk Oh, Jeonghun Sohn and Se Kwon Kim for fruitful discussions. 
}


\vspace{5mm}
\noindent{\textbf{Author contributions}}\\
{\small J.M.L. performed theoretical calculations and wrote the manuscript with the help from M.-J.H. and H.-W.L. All authors discussed the theoretical calculation and commented on the manuscript.
}

\vspace{5mm}
\noindent{\textbf{Competing interests}}\\
{\small The authors declare no competing interests.
}

\vspace{5mm}
\noindent{\textbf{Additional Information}}\\
{\small \textbf{Correspondence} and requests for materials should be addressed to M.-J.H. or H.-W.L.
}

\def\bibsection{\section*{\refname}} 

\bibliography{BibRef}

\end{document}